\documentclass[prb,showpacs,twocolumn,preprintnumbers,amsmath,amssymb,superscriptaddress]{revtex4}
\usepackage{graphicx}
\usepackage{color}
\usepackage{bm}
\usepackage{epstopdf}
\usepackage{times}

\newcommand{\ket}[1]{{\,\vert\, #1 \,\rangle}}

\begin{document}

\newcommand{\figurewidth}{2.8in}

\title{Determination of the phonon sidebands in the photoluminescence spectrum of semiconductor nanoclusters from \textit{ab initio} calculations}

\author{Peng Han}
\affiliation{Department of Physics, Beijing Key Lab for Metamaterials and Devices, Capital Normal University, Beijing 100048, China}
\affiliation{Department of Chemistry and Physics, University of Hamburg, Luruper Chaussee 149, Hamburg 22761, Germany}
\author{Gabriel Bester}
\email[E-mail:]{gabriel.bester@uni-hamburg.de}
\affiliation{Department of Chemistry and Physics, University of Hamburg, Luruper Chaussee 149, Hamburg 22761, Germany}
\affiliation{The Hamburg Centre for Ultrafast Imaging, Luruper Chaussee 149, D-22761 Hamburg, Germany}

\date{\today}
\begin{abstract}
	We propose a theoretical approach based on (constrained) density functional theory and the Franck-Condon approximation for the calculation of the temperature dependent photoluminescence of nanostructures.  The method is computationally advantageous and only slightly more demanding than a standard density functional theory calculation and includes transitions into multiphonon final states (higher class transitions). We use the approach for Si and CdSe colloidal nanoclusters (NCs) with up to 693 atoms and obtain very good agreement with experiment which allows us to identify specific peaks and explain their origin. Generally, breathing type modes are shown to dominate the phonon replicas, while optical modes have significant contributions for CdSe NCs and play a lesser role in Si NCs. We obtain significant anti-Stokes peak starting at 140K for Si NC explaining the broadening observed in the corresponding experiment.  We also apply the method to small molecular-like carbon structures (diamondoids), where electron-phonon coupling is typically large, and find that multiphonon processes (up to class 4) are very relevant and necessary to compare favorably with experiment. While it is crucial to include these multiphonon states in the small diamondoids with few tens of atoms, neglecting them in only marginally larger Si$_{87}$H$_{76}$ and Cd$_{43}$Se$_{44}$H$^*_{76}$ (and larger) quantum dots represents a good approximation.
\end{abstract}


\maketitle

\section{Introduction}

The main application fields of colloidal nanoclusters (NCs), such as
optoelectronics, quantum information processing, photovoltaics, light-emitting devices, or biolabeling\cite{yin05,talapin10,pandey08,salant10,meinardi14, bourzac13,achtstein14,shirasaki13}, rely on the quantum confined optical properties. In contrast to bulk materials, the quantum confinement effect encountered in NCs results in the emergence of discrete electronic as well as vibrational states. The confined nature of both the electronic and the vibrational states often results in a strong coupling referred to as vibronic coupling.  Photoluminescence spectroscopy (PLS) can be a powerful experimental tool to investigate this vibronic coupling\cite{dabbousi97,klimov07,chilla08,meulenberg09,huxter13}.
Indeed, in contrast to the sharp and singular peaks observed in atomic PLS, the strong coupling between electrons and the lattice vibration in NCs leads to the appearance of a multitude of phonon sidebands (satellites) leading to a rich and complex spectrum\cite{Yu10} that requires a solid theoretical interpretation. The vast knowledge available for bulk systems\cite{Yu10} or specific molecules is only marginally applicable to NCs, which represent a significantly different class of materials. Earlier investigations on vibronic coupling in NCs have shown the importance of phonon sidebands originating from longitudinal optical (LO)  modes and surface optical (SO) modes as well as torsional and spheroidal acoustic modes \cite{valerini05,lindwall07,chilla08,baker13,fernee14, pevere18}.

\begin{figure}
\centering
\includegraphics[width=0.4\textwidth]{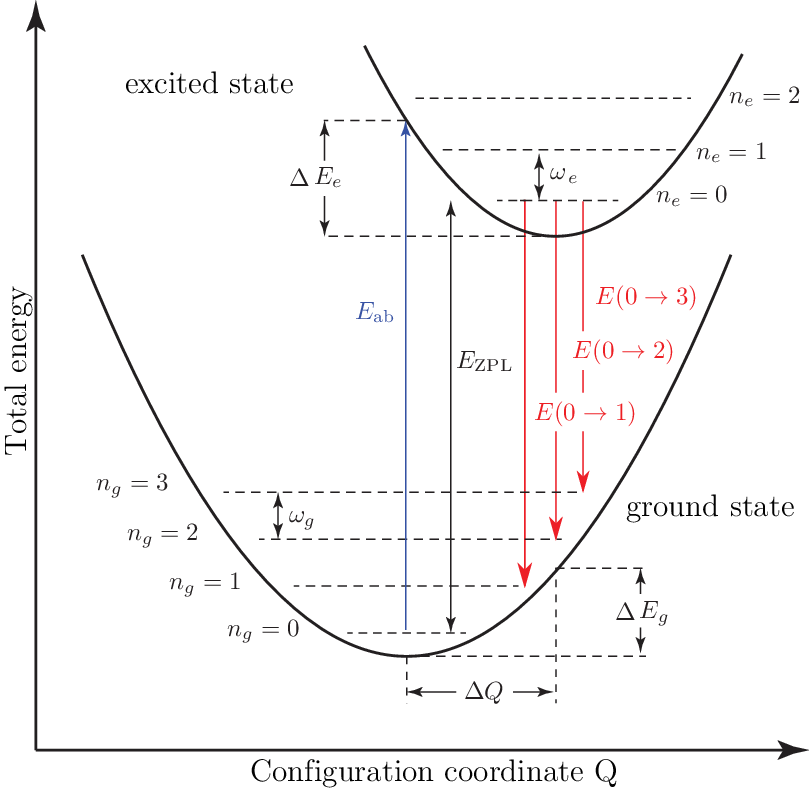}
\caption{\label{fig:FrankCondon} Schematic illustration of the configuration coordinate diagram describing the optical absorption and emission processes. The minima of the ground and excited state potential energy surfaces are displaced by $\Delta Q$. The relaxation, or reorganization, energies are shown as $\Delta E_{\{g,e\}}$ for the ground- and excited-state, respectively; $E_{\rm ZPL}$ indicates the zero-phonon line. Transitions from the lowest vibrational level of the excited electronic state with number of phonons $n_e=$ 0 to the ground electronic state with excited vibrational states, where the number of phonons $n_g=$ 1, 2, and 3, are labeled with red arrows.
}
\end{figure}

The common procedure to understand phonon satellites in the PLS of NCs is based on the Franck-Condon picture as schematically illustrated in Fig.~\ref{fig:FrankCondon}.
This approximation makes statements about the probabilities of the individual vibronic transitions, i.e., the transition between the dressed electron-vibron (vibronic) coupled state\cite{Haken04}. Within the Franck-Condon framework, the optical excitation is instantaneous (vertical transitions), so that the excited state has no time to relax into a more favorable configuration, which leaves the system in an excited vibronic state where not only the electronic part is excited (the state occupies the upper potential energy surface), but also in an excited state of vibration. The probability of the vibronic transition is thereby proportional to the overlap of the two nuclear wave functions. Using the Franck-Condon principle, phonon sidebands in the PLS of semiconductor NCs have been calculated by using both continuum models\cite{fomin98,lindwall07,chernikov12,callsen15} and \textit{ab initio} approaches\cite{gali16}. Although the exciton-phonon coupling strengths of NCs can be calculated using continuum models in the case of self-assembled quantum dots (QDs) with up to one hundred thousand atoms\cite{verzelen02}, the validity of such treatment for colloidal QDs with only hundreds to thousands of atoms is questionable. Moreover, in most of the empirical model calculations, only contributions from bulk-like LO phonon modes are considered. In contrast to empirical methods, \textit{ab initio} based studies on PLS have focused on small molecules, due to the present computational limitations of state-of-the-art first-principles approaches. \textit{Ab initio} studies of NCs, with the aspiration of a deeper understanding on vibronic coupling in quantum confined systems, are still needed.

In this work, we present a theoretical \textit{ab initio} approach suitable to study the exciton-phonon interaction and their effects on phonon satellites in the PLS of colloidal Si and CdSe NCs with radii ranging from 7.8 to 14.9~\AA~ and five different small carbon clusters called diamondoids. Using density functional theory (DFT) and constrained density functional theory (CDFT) approaches, we calculate the nuclear coordinates of colloidal nanostructures with and without photoexcitation and the vibrational modes of the ground state. The Huang-Rhys (HR) factors of each vibrational mode are further calculated using the procedure given in Refs.~\onlinecite{alkauskas12} and~\onlinecite{han19}. Within the Franck-Condon principle and using the harmonic oscillator approximation for the nuclear wave functions\cite{Grosso15} we are able to calculate the line shapes of PLS at both zero and finite temperatures. We extend the approach to calculate the effects of multiphonon processes where different types of vibrons participate simultaneously. These processes are classified into classes. We find that for our larger CdSe and Si NCs a treatment including only class 1 transitions represents a good approximation, while for our smallest diamondoid structure calculations must encompass class 4 processes to accurately reproduce experiment.  

\section{Theoretical Approach}
\label{sec:theory}

% Structure relax
In the first step, we optimize the geometry of the different structures. The equilibrium positions of the nuclei in the ground and the excited states of spherical quantum dots are optimized under constrained symmetry, while the structures of the excited diamondoids are optimized without symmetry constraint until the forces are reduced to less than 3$\times$10$^{-6}$ Hartree/Bohr by using DFT and CDFT implemented in the CPMD code\cite{cpmd08}. In order to avoid interactions with periodic images of the structure, the NCs are located in the center of a simple cubic supercell with more than 5~\AA~ separation between the outermost atoms and the closest boundary.
% CDFT
In CDFT, electronic states and interatomic forces of excited NCs are calculated by constraining one electron to occupy the lowest unoccupied molecular orbital (LUMO) leaving one hole at the highest occupied molecular orbital (HOMO) via the restricted open shell Kohn-Sham algorithm\cite{dederichs84,qiu06,frank98}. 
% LDA etc..
The local density approximation (LDA) and Trouiller-Martin norm-conserving pseudopotentials are used in both DFT and CDFT calculations with energy cutoff of 60, 40, and 50 Ry for diamondoids, Si, and CdSe NCs, respectively. 

% Phonons
Once the geometry has been optimized, we calculate the dynamical matrix elements via finite difference with the self-consistently calculated electronic states\cite{Yu10, han12a}. The vibrational eigenmodes along with the corresponding eigenvectors are then obtained by solving the dynamical matrix equation,
\begin{equation}
\label{eq:eigen}
\sum_{J}\frac{1}{\sqrt{M_{I}M_{J}}}\frac{\partial^{2}V({\bm R})}{\partial{\bm R^{I}}\partial{\bm R^{J}}}
{\bm X^{J}}=\omega^{2}{\bm X^{I}},
\end{equation}
where $I$ and $J$ label the atoms, $M$ denotes the atomic masses, $V({\bm R})$ is the potential energy, ${\bm R}$ represents the atomic position, and ${\bm X}$ is the vibrational eigenvector with the vibrational frequency $\omega$.

Using the ground $\{\bm R^{I}_g\}$ and the excited state $\{\bm R^{I}_e\}$ atom positions we calculated the structural rearrangement in the configuration coordinates as\cite{Shuai12,alkauskas12}
\begin{equation}
\label{eq:dq}
\Delta Q_{i} = \sum_{I}\sqrt{M_{I}}(\bm{R}_e^{I}-\bm{R}_g^{I})\cdot\bm{X}_{i}^{I},
\end{equation}
where $\bm X^{I}_i$ is the unit vector of atom $I$ in the direction of the normal vibrational mode $i$ in the ground state and $\bm{R}_e^{I}-\bm{R}_g^{I}$ denotes the lattice distortion between the excited and the ground states. The HR factor of vibrational mode $i$ is thereafter calculated as\cite{huang50,merlin78}
\begin{equation}
\label{eq:hrfactor}
S_i=\frac{1}{2\hbar}\omega_i\left(\Delta Q_{i} \right)^2.
\end{equation}

The PLS intensity of a photoexcited NC can be calculated via Fermi's golden rule\cite{Grosso15} as
\begin{equation}
\label{eq:Iem}
I_{em}(E)=\frac{2\pi}{\hbar}\sum_{n}|\langle\psi_e\phi_m|{\bm \mu}|\psi_g\phi_{n}\rangle|^2\delta(E_{em}-E_{gn}-E)  ,
\end{equation}
where $|\psi_g\rangle$ and $|\psi_e\rangle$ are the ground and the excited state electronic wave functions, $|\phi_{m}\rangle$ and $|\phi_{n}\rangle$ denote the nuclear (also called ionic)  wave functions, $\bm{\mu}$ is the dipole operator for electronic transitions, $E$ is the energy of the emitted photon, $E_{em}$ are the energies of the excited electronic system with nuclear wavefunction $\phi_m$ and  $E_{gn}$ are the energies of the ground electronic system with nuclear wave function  $\phi_n$.

The Frank-Condon approximation significantly simplifies the problem by assuming a separation of electronic and vibrational wavefunctions:
\begin{equation}
\label{eq:FCapprox}
I_{em}(E)=\frac{2\pi}{\hbar}|\langle\psi_e|{\bm \mu}|\psi_g\rangle|^2  \sum_{n}|\langle\phi_m | \phi_{n}\rangle|^2\delta(E_{em}-E_{gn}-E)  .
\end{equation}
In this approach the electronic transition dipole moment is assumed to be independent of the nuclear coordinates. An improved description is given by the Herzberg-Teller approach, \cite{herzberg33} where the electronic transition integrals are expanded in powers of the nuclear displacement. This approach, which is not followed in this work, leads to improved results especially for optical transitions that are spin or orbitally forbidden and consequently dark according to the Frank-Condon formulation.

The nuclear wave function in the ground and excited electronic states can be generally written as $|\phi_n\rangle = |n_g^1, n_g^2, n_g^3, ...n_g^i, ... n_g^{3N-6}\rangle$ and $|\phi_m\rangle = |n_e^1, n_e^2, n_e^3, ...n_e^i, ... n_e^{3N-6}\rangle$ with $n_g^i (n_e^i)$ the vibron occupation number of the vibrational mode $i$ in the ground (excited) electronic state. The superscript $i$ on $n_g^i$ defines the vibrational mode and runs from 1 to $3N-6$. The ground state vibration in this notation is given as $\phi_0 = |0,0,0,...,0 \rangle$.

\begin{figure}[t]
%\centering
	\includegraphics[width=0.4\textwidth]{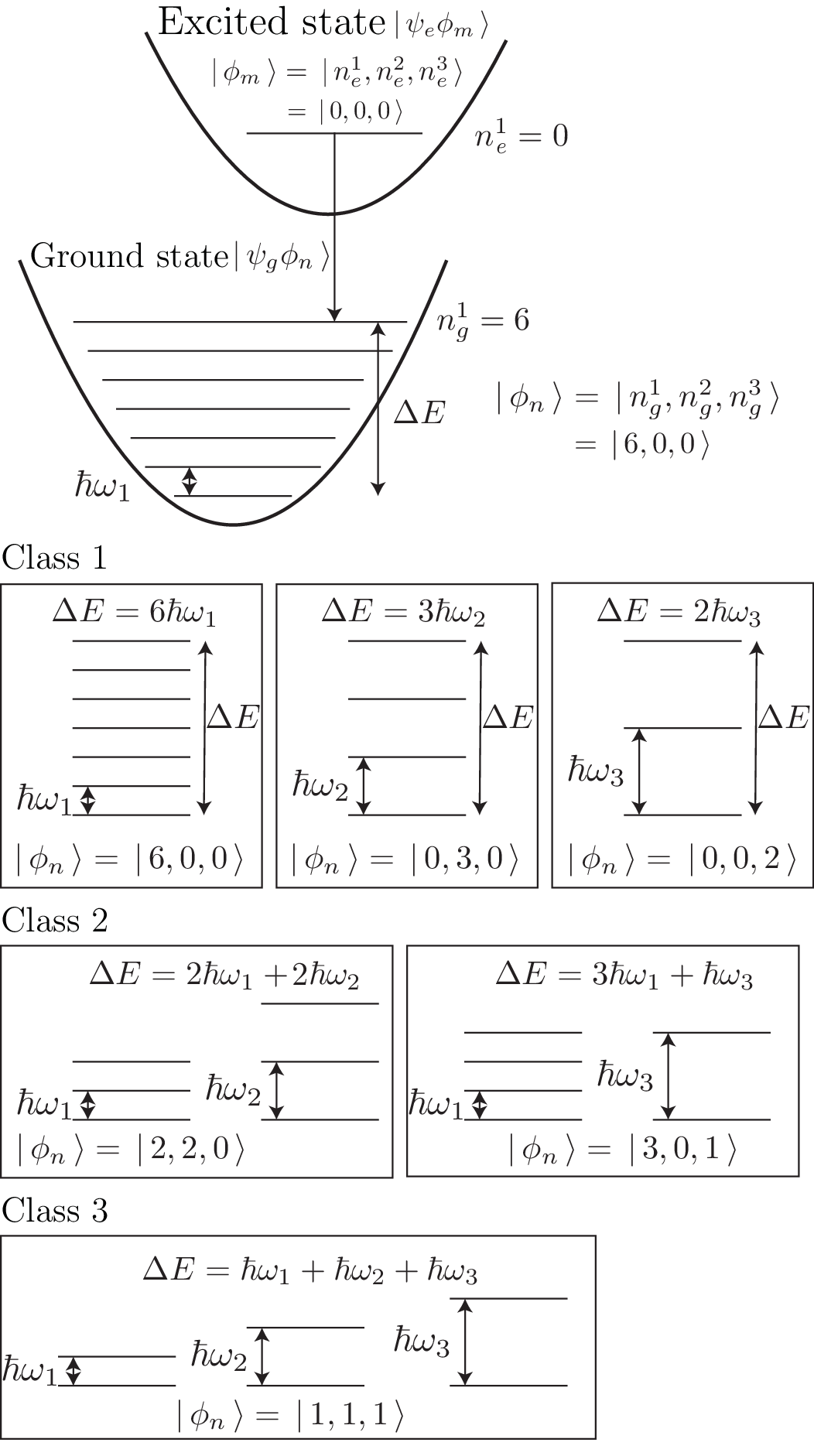}
\caption{\label{fig:PL_classes} Schematic illustration of vibronic transitions and their classification into classes. The different nuclear wave functions depicted all correspond to an energy of 6$\hbar\omega_1$. One class 2 transition with the same energy has been left out ($\ket{4,1,0}$) for space reasons.}
\end{figure}

In Fig.~\ref{fig:PL_classes}, we schematically illustrate a vibronic transition in a toy model system with three vibrational modes with frequencies $\omega_{1,2,3}$. As shown in the top of Fig.~\ref{fig:PL_classes}, the initial vibronic state $|\psi_e\phi_m \rangle$ is composed of the excited electronic state $\ket{\psi_e}$ and the ground vibrational state $\ket{\phi_m} = \ket{n_e^1, n_e^2, n_e^3} = \ket{0,0,0}$ (this corresponds to zero vibrational temperature), while the final state $\ket{\psi_g \phi_n}$ is composed of the ground electronic state $|\psi_g\rangle$ and the vibrational state $\ket{\phi_n} = \ket{n_g^1, n_g^2, n_g^3}$.
By assuming a reorganization energy of $\Delta E = 6\hbar\omega_1 = 3\hbar\omega_2 = 2\hbar\omega_1$, we have seven different configurations of the final vibrational state $|\phi_n\rangle$. In Fig.~\ref{fig:PL_classes} we show only six of them for space reasons, the missing one being the nuclear wave function $\ket{4,1,0}$. These vibrational state configurations can be divided into class 1, class 2, and class 3 according to the number of vibrational modes simultaneously excited in the vibronic transition. In the case of class 1, the configurations $|\phi_n \rangle =|6, 0, 0 \rangle$, $|\phi_n \rangle =|0, 3, 0 \rangle$, and $|\phi_n \rangle =|0, 0, 2 \rangle$ correspond to the excitation of six vibrons with $\omega_1$ frequency or three vibrons with $\omega_2$ frequency or two vibrons with $\omega_3$ frequency. In class two and class three, different vibrational modes can be excited simultaneously.

Direct calculation of the overlap of nuclear wave functions $\langle \phi_m|\phi_n \rangle$ requires highly multidimensional integrals with the normal modes of the excited and ground electronic states $Q_g$ and $Q_e$  via the Duschinsky transformation\cite{duschinsky37,borrelli03,santoro07,barone09,banerjee14}. This algorithm is challenging to nanostructures with a large number of atoms due to its very large computational requirements. In order to circumvent this difficulty, a simplified approach based on the so-called linear-mode approximation or parallel-mode approximation is used to calculate the overlap of the nuclear part of the wave functions in nanostructures or defect structures in solid\cite{lax52,keil65,alkauskas12}.
Within this approximation, the overlap between the nuclear (or ionic) wave functions is analytically calculated from the overlap between displaced harmonic oscillator wave functions as\cite{keil65,Grosso15,gali16}
\begin{equation}
\label{eq:hr}
	|\langle\phi_{m}|\phi_{n}\rangle|^2=\prod_{i=1}^{3N-6} |\langle n_e^i|n_g^i\rangle|^2,
\end{equation}
with
\begin{equation}
\label{eq:hr_mode}
	|\langle n_e^i | n_g^i\rangle|^2 = \text{e}^{-S_i}\frac{n_e^i!}{n_g^i!}{S_i^{n_g^i-n_e^i}}|L_{n_e^i}^{n_g^i-n_e^i}(S_i)|^2 \quad .
\end{equation}
where $L_{n_e^i}^{(n_g^i-n_e^i)}(S_i)$ denotes the generalized Laguerre polynomials with $n_g^i$ and $n_e^i$ the number of vibrons of mode $i$ in the ground and excited electronic states, respectively. The superscript on the HR factor signifies the power of the difference between $n_g^i$ and $n_e^i$.
Combining Eq.~(\ref{eq:Iem}) with Eqs.~(\ref{eq:hr}) and~(\ref{eq:hr_mode}), the intensity of the PLS is given as
\begin{eqnarray}
\label{eq:Iem2}	
	&I_{em}(E) &= \frac{2\pi}{\hbar}|\bm{\mu}_{eg}|^2 \nonumber \\
	&&
	\times \sum_{n_g^i=0}^\infty 
\left( 
\prod_{i=1}^{3N-6}\text{e}^{-S_i}\frac{n_e^i!}{n_g^i!}{S_i^{n_g^i-n_e^i}}|L_{n_e^i}^{(n_g^i-n_e^i)}(S_i)|^2
\right) \nonumber \\
	&&
\times \delta[E_0-E-\sum_i(n_g^i-n_e^i)\hbar\omega_i],
\end{eqnarray}
where $n_g^i$ under the sum sign is a compact notation for a sum over all possible final vibrational states, i.e., combination
of the $n_g^i$ indices.  $\bm{\mu}_{eg}=\langle\psi_e|\bm{\mu}|\psi_g\rangle$ represents the dipole matrix elements. 
The frequency mismatch between ground and excited electronic states is ignored in our approach and we use vibrational frequencies of only the ground state. We will show differences between vibrational frequencies in the ground and excited states in Sec.~\ref{sec:HighClass}.

The values of $n_e^i$ are determined from the initial state and will depend on temperature.  At zero temperature, for instance, $n_e^i$ is zero for all vibrational modes (as we used for the initial state in Fig.~\ref{fig:PL_classes}). 
We therefore assume a perfect exciton thermalization. This can be justified by the different timescales: the exciton radiative lifetime is around hundreds of nanoseconds for CdSe QDs~\cite{donega06} to a few microseconds for Si QDs~\cite{sangghaleh13}; the thermalization process via phonon emission is typically in the picosecond range~\cite{zhang21}.  With increasing temperature the vibron states are becoming populated and $n_e^i$ may be larger than zero (we will later use a Boltzmann distribution to take this into account). 
The energy conservation is given by the delta function where we used $E_0 = E_{e0}-E_{g0}$ defining the energy difference between the excited and the ground electronic states in the absence of vibrons.

Since $\bm{\mu}_{eg}$ is independent on lattice vibrations, we will focus on the remaining contributions originating from the nuclear part of the vibronic wave function. We will consider $A_{em}(E)$, where we have used a Lorentz broadening for each phonon sideband peak:
\begin{eqnarray}
\label{eq:Iem3}
	&I_{em}(E) &\propto A_{em}(E) \nonumber \\
	&& =\sum_{ n_g^i=0}^\infty   %\nonumber \\
\left(
\prod_{i=1}^{3N-6}\text{e}^{-S_i}\frac{n_e^i!}{n_g^i!}{S_i^{n_g^i-n_e^i}}  |L_{n_e^i}^{(n_g^i-n_e^i)}(S_i)|^2
\right) \nonumber \\
	&& 
	\times
\frac{\gamma^2}{[E_0-E-\sum_i(n_g^i-n_e^i)\hbar\omega_i]^2+\gamma^2},
\end{eqnarray}
where the broadening of the Lorentz function $\gamma$ is taken as the inverse of the phonon lifetime (which is not calculated in this paper but estimated from experiment).

\section{Low temperature phonon sidebands for $Si$ and $CdSe$ NCs}

% General Section on Si and CdSe
The Si and CdSe NCs are constructed by cutting a sphere out of bulk diamond (Si) or zinc-blende (CdSe) centering the structure on a silicon (cadmium) atom. Surface atoms with only one nearest-neighbor bond are subsequently removed. This procedure leads to a $T_d$ point group symmetry in both cases. To avoid defect states the surface atoms are passivated by hydrogen atoms (for Si NCs) or pseudohydrogen atoms (for CdSe NCs) with a fractional charge of 1/2 and 3/2 for the passivation of Se and Cd atoms, respectively. 

\begin{figure}[t]
\centering
\includegraphics[width=0.4\textwidth]{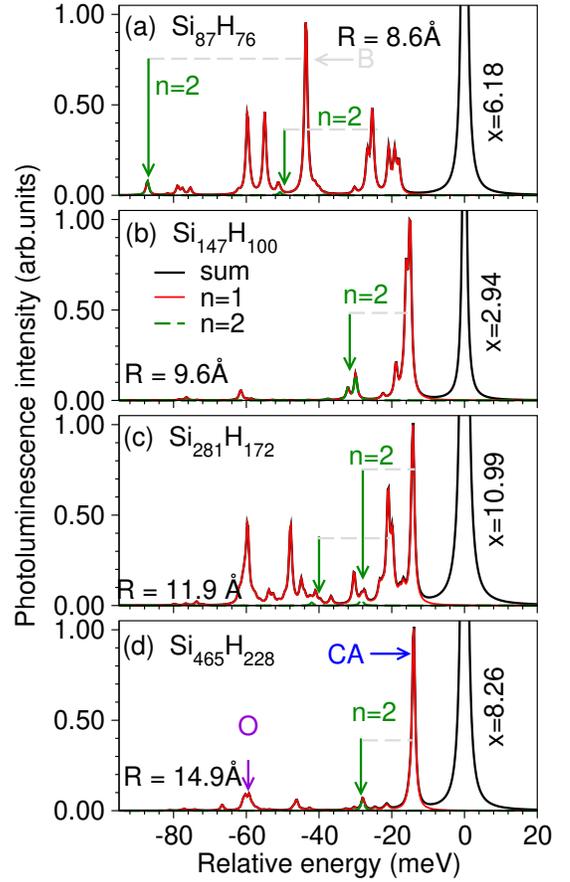}
\caption{\label{fig:Si_PL}
	Calculated zero-temperature PLS intensity (black line) of Si NCs from Eq.~(\ref{eq:Iem3}) with contributions of phonon number 
	$n_g = 1$ (red), $n_g = 2$ (green) and the sum 0 to 10 (black),  for (a) Si$_{87}$H$_{76}$, (b) Si$_{147}$H$_{100}$, (c) Si$_{281}$H$_{172}$, and (d) Si$_{465}$H$_{228}$ NCs. The intensity of the strongest phonon satellite peak is set to 1.0 and the intensity of the ZPL peak is given as ``x= ''. $E_{ZPL}$ is set to zero. ``CA'' indicates the coherent acoustic mode. ``O'' and the gray bar indicates modes with optical character, the black bar within the gray area shows the energy of the LO bulk mode. ``B'' indicates a breathing type mode described in the text.}
\end{figure}

\begin{figure}
\centering
\includegraphics[width=0.4\textwidth]{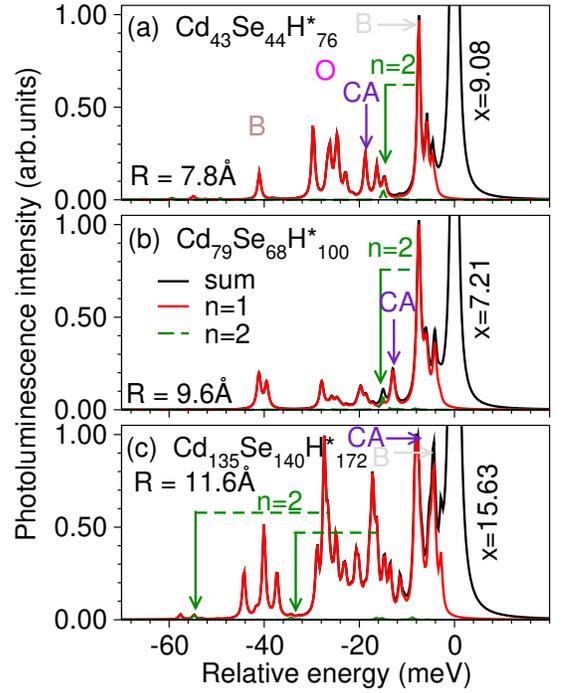}
\caption{\label{fig:CdSe_PL}
	Idem to Fig.~\ref{fig:Si_PL} but for (a) Cd$_{43}$Se$_{44}$H$^*_{76}$, (b) Cd$_{79}$Se$_{68}$H$^*_{100}$, and (c) Cd$_{135}$Se$_{140}$H$^*_{172}$ NCs. ``S" indicates surface type modes.  
}\end{figure}

In Fig.~\ref{fig:Si_PL} and Fig.~\ref{fig:CdSe_PL}, we  plot the calculated PLS of different Si and CdSe NCs as a function of the relative energy to the ZPL.
The PLS are calculated  at zero temperature, i.e., the vibron occupation number in the excited state is zero ($n_e^i =  0$). The results are broadened with a Lorentz function with a width of 1~meV, which corresponds to the phonon lifetime of the bulk material. The sum over vibron occupation numbers $n_g^i$ is taken from 0 to 10 (black lines) for $n_g^i = $ 1 (red lines) and $n_g^i = $ 2 (green lines).

To compare the intensity of the different phonon satellites, we set in Fig.~\ref{fig:Si_PL} and ~\ref{fig:CdSe_PL} the intensity of the strongest satellite peak to $1.0$ and give the intensity of the ZPL peak in the figure as numerical value (``x='').
We see from Fig.~\ref{fig:Si_PL} and Fig.~\ref{fig:CdSe_PL}  that the amplitude of the ZPL peak (transitions from $n_e^i=0$ to $n_g^i=0$, black line) is much stronger than those of the phonon satellites and most of the phonon satellites are dominated by vibronic transitions with $n_e^i = 0 \rightarrow n_g^i = 1$ (red lines much larger than green lines).

To rationalize these results, we rewrite Eqs.~(\ref{eq:hr}) and ~(\ref{eq:hr_mode}) at zero temperature with $n_e^i = 0$ for all vibrations $i$. The square of the overlap of nuclear wave functions can be simplified in this case as
\begin{equation}
\label{eq:hr_0T}
	|\langle\phi_{0}|\phi_{n}\rangle|^2=\prod_{i=1}^{3N-6}\frac{S_i^{n_g^i}}{n_g^i!}\text{e}^{-S_i}\quad ,
\end{equation}	
where $S_i^{n_g^i}$ means  $S_i$ to the power of ${n_g^i}$ and the index $n$ is defined from $|\phi_n\rangle = |n_g^1, n_g^2, n_g^3, ...n_g^i, ... n_g^{3N-6}\rangle$.
The ZPL is accordingly given as $|\langle\phi_{0}|\phi_{0}\rangle|^2 = \prod_i^{3N-6} \text{e}^{-S_i}$
which is much larger  than  Eq.~(\ref{eq:hr_0T}) when  ${n_g^i}>0$ and $S<1$.
Moreover, as indicated in Eq.~(\ref{eq:hr_0T}), the value of $|\langle\phi_{0}|\phi_{n}\rangle|^2$ decreases rapidly with an increasing number of vibrons $n$ for small HR factor ($S_i < 1$). This results in the prevalence of phonon sidebands with phonon number $n_g^i=1$ in the PLS (red lines in Fig.~\ref{fig:Si_PL}).
This is in agreement with our general understanding that weak vibronic coupling (small $S_i$) leads to a small displacement of the electronic charge densities going from the ground to the electronically excited state, and therefore involves nuclear motion to a lesser extent.

% Size dependence
From Figs.~\ref{fig:Si_PL} and ~\ref{fig:CdSe_PL} it is surprisingly difficult to identify a clear size dependence in the intensity of the phonon side bands. Generally, the spectra show a very large size/structure dependence, but extracting trends is difficult.  For Si NC (Fig.~\ref{fig:Si_PL}), comparing our two largest structures (where the structures become more bulk-like and less molecular) we could conclude that the strongest peak at 15 meV (that we will later identify as coherent acoustic mode) seems to remain strong and size independent, while the other phonon satellites become much weaker. Also for CdSe NCs (Fig.~\ref{fig:CdSe_PL}), if we discard the very small Cd$_{43}$Se$_{44}$H$^*_{76}$ we can conclude that the intensity of the phonon replicas decreases with increasing cluster size. This latter observation is in agreement with the idea that exciton-induced lattice distortions become smaller with increasing NC size leading to a decreasing HR factor.

% Specific Silicon

% Coherent acoustic mode
Next, we focus on the Si NCs (Fig.~\ref{fig:Si_PL}), where we see the strongest phonon satellite at around 15~meV [indicated with CA in Fig.~\ref{fig:Si_PL}~(d)] and the corresponding two-phonon replicas (in green) at around 30~meV from the ZPL. This dominant satellite  exhibits a red shift with increasing cluster size. By looking at their eigenmodes we  identify them as coherent acoustic phonon modes\cite{han12a} (or ``breathing modes'' with an in phase atomic vibration). As a vibrational mode with $\Gamma_1$ point group symmetry, the coherent acoustic phonon mode strongly couples with the lattice distortion induced by photon excitation and results in a strong phonon replica in PLS. Moreover, the confinement effect leads to the red shift of the coherent acoustic phonon with increasing cluster size from 19.2~meV in Si$_{87}$H$_{76}$ NC (a) to 15.0~meV in Si$_{147}$H$_{100}$ NC (b), 14.6~meV in Si$_{281}$H$_{172}$ NC (c), and 14.0~meV in Si$_{465}$H$_{228}$ NC (d), which have been accurately described by using the classical Lamb model\cite{lamb82,han12a}.

% Bulk LOTO mode
In addition to the coherent acoustic phonon modes, another interesting phonon can be seen at around 60~meV [indicated with O in Fig.~\ref{fig:Si_PL}~(d)]. This mode is an optical-like phonon mode, close in frequency to the LO mode in bulk silicon at 63~meV,  and the frequency of such mode is nearly constant over the entire cluster size range studied.

% Another type of breathing mode
The dominant peak at 43~meV in Si$_{87}$H$_{76}$ [indicated with a B in Fig.~\ref{fig:Si_PL}~(a)] is induced by another type of coherent acoustic phonon mode, where the vibration can be separated into a breathing core part and a breathing shell part with opposite phases. 

% Specific CdSe results

% Acoustic
The results for CdSe NCs (Fig.~\ref{fig:CdSe_PL}) show a phonon satellite peak at 18.7~meV in Cd$_{43}$Se$_{44}$H$^*_{76}$ NC (a), 12.9~meV in Cd$_{79}$Se$_{68}$H$^*_{100}$ (b), and 9.2~meV in Cd$_{135}$Se$_{140}$H$^*_{172}$ NC (c) that we can identify as a coherent acoustic mode (breathing mode, indicated as CA in Fig.~\ref{fig:CdSe_PL}). The strongest peaks at 7.6~meV of Cd$_{43}$Se$_{44}$H$^*_{76}$ NC (a), 7.5~meV in Cd$_{79}$Se$_{68}$H$^*_{100}$ (b), and 4.5~meV in Cd$_{135}$Se$_{140}$H$^*_{172}$ NC (c) (indicated as B in Fig.~\ref{fig:CdSe_PL}) originate from a type of breathing mode where the structure oscillates between oblate and prolate shapes.  As a confined mode, we see a red shift of these peaks with increasing cluster size.

% Optic
The bulk CdSe LO phonon is at 26.4~meV and we can identify the optical character of the NC phonon modes at this energy in all three structures. The contribution from optical modes seems to become larger for larger structures. In the case of  Cd$_{135}$Se$_{140}$H$^*_{172}$ the optical modes is as strong as the acoustic modes.
In comparison to the silicon NC case, the phonon replica originating from the LO phonon of bulk CdSe  play a more important role in the calculated PL spectrum of large CdSe NC as a result of the stronger LO phonon coupling in ionic materials.

%%%%%%%%%%%%%%%%%%%%%%%%%%%%%%%%%%%%%%%%
\section{Temperature dependence of phonon sidebands}

At elevated temperatures, the thermally excited vibronic states of the initial state must be included in the calculation, which means that we need to calculate the nuclear wave function $\ket{\phi_m}$, i.e., the occupation numbers $n_e^i$ of the vibron state.
For this purpose we use a Boltzmann distribution for the occupation of the excited state as
\begin{equation}
	f(n_e^i,T) = 
	\frac{\text{e}^{-\sum_i^{3N-6} { n_e^i\hbar\omega_i}/{k_\text{B}T} }} 
	{\sum_{ n_e^{i}=0}^\infty \text{e}^{-\sum_i^{3N-6}{n_e^i\hbar\omega_i}/{k_\text{B}T} }} \quad, 
\end{equation}
where $n_e^i\hbar\omega_i$ is the energy of the vibration with occupation number $n_e^i$.
The temperature-dependent PLS is therefore calculated as\cite{ulstrup75},
\begin{eqnarray}
\label{eq:Iem3_T}
	&A_{em}(E,T)&=\sum_{ n_e^i=0}^\infty \sum_{ n_g^i=0}^\infty f(n_e^i,T)\nonumber \\
	&& \times
\left(
         \prod_{i=1}^{3N-6} \text{e}^{-S_i}\frac{n_e^i!}{n_g^i!}{S_i^{n_g^i-n_e^i}}|L_{n_e^i}^{(n_g^i-n_e^i)}(S_i)|^2
\right) \nonumber \\
	&& \times
         \frac{\gamma^2}{[E_0-E-\sum_i(n_g^i-n_e^i)\hbar\omega_i]^2+\gamma^2}.
\end{eqnarray}

By increasing temperature, the occupation number of the excited vibron states becomes non-zero leading to the vibronic transitions between $n_e^i > 0$ and $n_g^i = 0$ states. These transitions result in anti-Stokes shift in the PLS and the possibility of such transitions increases rapidly with increasing temperature.

\begin{figure}
\centering
\includegraphics[width=0.45\textwidth]{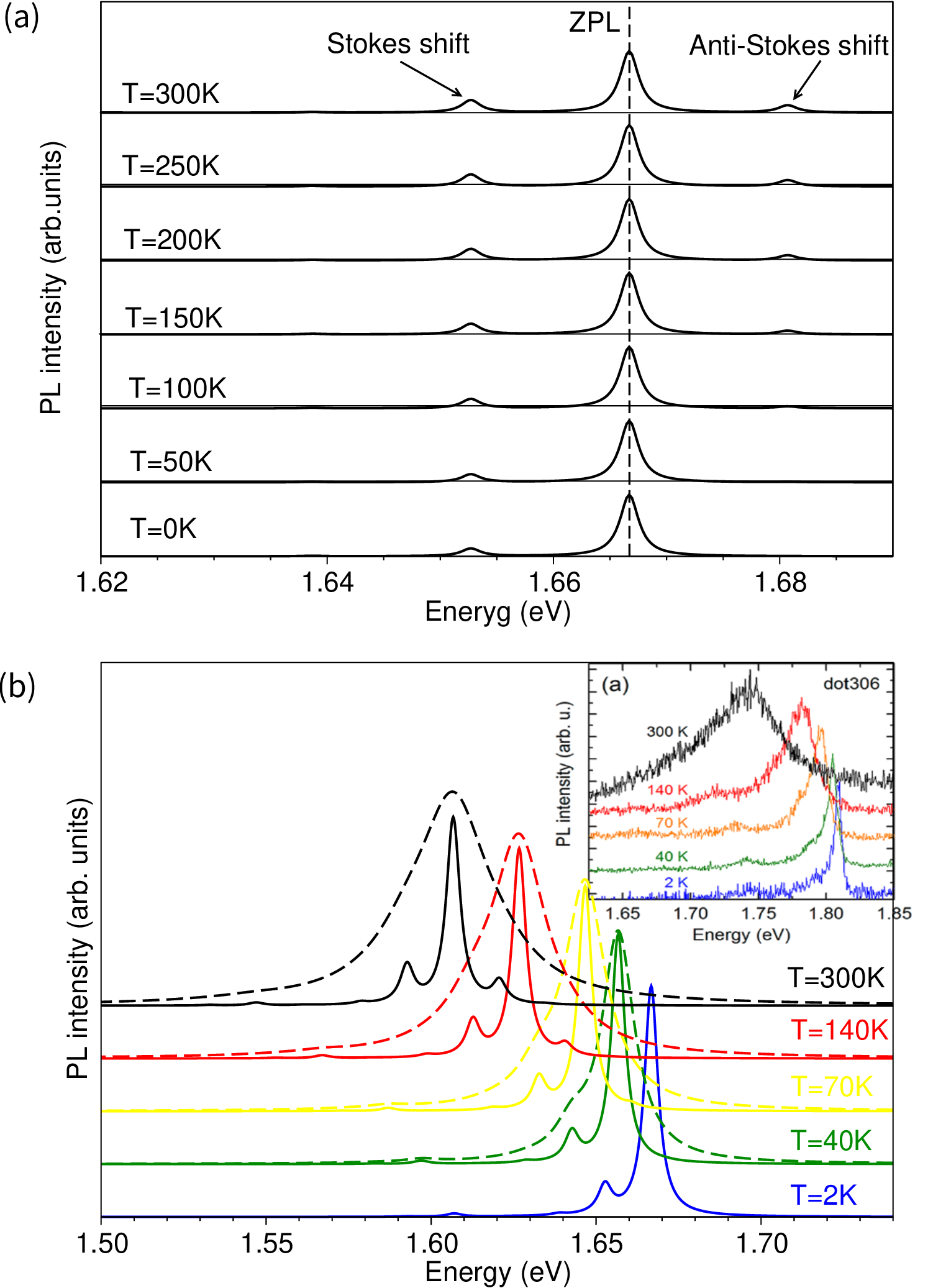} 
\caption{\label{fig:Si_temperature}
(a) Calculated PLS of Si$_{465}$H$_{228}$ NC at different temperatures focusing on the energy region of the coherent acoustic mode.
(b) Calculated PLS of Si$_{465}$H$_{228}$ NC at different temperatures corresponding to experiment [\onlinecite{pevere18}] using a Lorentz broadening of 5~meV (solid lines) and using a broadening of 12~meV, 15~meV, 20~meV and 25~meV (dashed lines). The results are shifted energetically according to the experiment (see text). 
Inset: measured normalized PLS of oxide-passivated Si-QD at different temperatures taken from Ref.~[\onlinecite{pevere18}].
}\end{figure}

In Fig.~\ref{fig:Si_temperature} (a), we plot the calculated PLS of a Si$_{465}$H$_{228}$ NC at temperatures of $T=$ 0, 50, 100, 150, 200, 250, and 300~K with a Lorentz broadening of 2~meV.
We notice that the intensity of the anti-Stokes peak increases with increasing temperature, especially at $T$ higher than 150~K. This is the result of the increased occupation of the excited vibronic states of the excited state ($n_e^i = 1$). We also notice that the intensity of the Stokes peak increases with temperature and attribute it to the appearance of vibronic transition between $n_e^i = 1$ to $n_g^i = 2$ (in addition to the $n_e^i = 0$ to $n_g^i = 1$ low temperature transition).

To compare with experiment, we replot our results in Fig.~\ref{fig:Si_temperature}(b) using the experimentally reported temperatures of 2~K (solid blue lines), 40~K (solid green lines), 70~K (solid orange lines), 140~K (solid red lines), and 300~K (solid black lines). We use a Lorentz broadening of 5~meV. The experimental results of Pevere \emph{et~al.}~\cite{pevere18} are shown as an inset.
Due to the thermal expansion of the lattice and mainly due to the electron-vibron zero point motion effect, the ZPL exhibits a red shift with increasing temperature\cite{doherty14,han13}. These effects are not calculated in this work but, in order to compare with experiments, the ZPL is shifted using the temperature shift taken from experiment \cite{pevere18}. 
We see the two strong phonon replicas in Fig.~\ref{fig:Si_temperature}(b) with energies around 15~meV and 60~meV we already described in Fig.~\ref{fig:Si_PL} as coherent acoustic phonon mode and optical-like phonon mode. As illustrated in the inset of Fig.~\ref{fig:Si_temperature} (b), these two characteristic phonon replicas are also revealed in the measured PLS of oxide-passivated Si QDs. By comparing the calculated PLS at various temperatures, we see an anti-Stokes peak appearing at temperatures higher than 140~K. Due to the increasing occupation of the excited vibronic states $n_e>0$ with increasing temperature, the intensity of both Stokes peaks (vibronic transitions between $n_e = 0$ and $n_g = 1$ and between $n_e = 1$ and $n_g = 2$) and anti-Stokes peaks (vibronic transition between $n_e = 1$ and $n_g = 0$) increase. At room temperature, the intensity of the anti-Stokes and Stokes peaks are comparable. This leads to a different line shape of the PLS at high temperature with a broad shoulder at higher energy. 
To facilitate the comparison with experiment~\cite{pevere18} (inset), we further broaden our theoretical results with a Lorentz broadening of 12, 15, 20, and 25~meV for the case of $T=$ 40, 70, 140, and 300~K, respectively, and show the results as dashed lines. 
The applied broadening is accounting for pure dephasing\cite{Scully97}.
The underlying process is thereby exciton-phonon scattering and depends on the phonon lifetime, which is decreasing with increasing temperature. The shorter phonon lifetime leads to a larger applied broadening. However, these lifetimes are not calculated in this work so that the broadening are simply applied according to the experimental results.
Using this broadening we obtain a very good agreement to the measured PLS.

%%%%%%%%%%%%%%%%%%%%%%%%%%%%%%%%%%%%%%%%%%
\section{Comparison with low-temperature experiment}
%%%%%%%%%%%%%%%%%%%%%%%%%%%%%%%%%%%%%%%%%%

\begin{figure}
\centering
\includegraphics[width=0.45\textwidth]{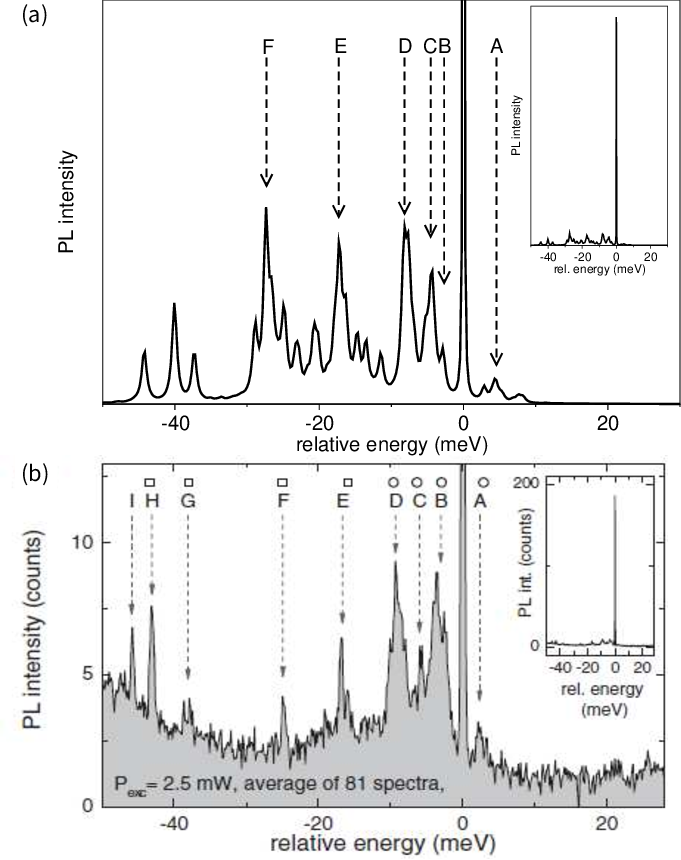}
\caption{\label{fig:CdSe_PL_exp}
(a) Calculated PLS of CdSe NC with $R=$11.6~\AA~ radius. The $E_{ZPL}$ has been set to zero and the temperature to 30K. Distinct phonon satellite peaks are marked by arrows and labeled A--F. (b) Measured photoluminescence spectrum of CdSe-CdS-ZnS core-shell-shell NC with core radius of 18~\AA~ and overall radius of about 23.5~\AA~. The experimental data is taken from Ref.~[\onlinecite{chilla08}].
}\end{figure}

In Fig.~\ref{fig:CdSe_PL_exp} we compare our calculated PLS of Cd$_{135}$Se$_{140}$H$^*_{172}$ cluster with 11.6~\AA~ radius to the measured PLS of CdSe-CdS-ZnS core-shell-shell NC with 18~\AA~ radius core and 23.5~\AA~ overall radius, taken from Ref.~[\onlinecite{chilla08}]. To avoid thermal broadening of the PLS, the measurements in Ref.~[\onlinecite{chilla08}] were performed at 8~K. The calculated PLS shown here corresponds to the structure already discussed in Fig.~\ref{fig:CdSe_PL}(c) but has been calculated at a temperature of 30~K using Eq.~(\ref{eq:Iem3_T}). Both the calculated and measured spectra are energetically shifted by setting the energy of the ZPL to zero. In Ref.~[\onlinecite{chilla08}], the distinct peaks measured in the PL spectrum were marked as A--I and we have followed this description for the theoretical results. 
% General good agreement
Comparing Figs.~\ref{fig:CdSe_PL_exp} (a) and \ref{fig:CdSe_PL_exp} (b), we see that we obtain a general good agreement and are able to find a correspondence between the peaks identified in the experiment and the theory.

% A-Peak
Among the phonon sidebands, peak A is an anti-Stokes peak induced by the vibronic transition between $n_e=1$ in the excited state and $n_g=0$ in the electronic ground state. Due to the weak phonon occupation of the excited electronic state at low temperature, we see the intensity of peak A is lower than that of the Stokes peaks. We obtain a good fit to the experiment for this peak using a temperature of 20 to 30 K, which is higher than the 8 K reported in the experiment. At a temperature of 8 K the vibronic state $n_e=1$ is almost unoccupied and peak A nearly absent. 
%\gbnote
{We can think about two different origins for this discrepancy. First, our simple assumption of a thermalized system (Boltzmann occupation of the excitonic states) may not correspond to the experimental situation and a more complete theoretical model involving excited state carrier dynamics may be required. In view of the dense manifold of vibrational states and the small separation of excitonic levels we, however, do not expect a particularly slow thermalization process~\cite{han15}. Second, the experimental conditions of the probed sample may not correspond to 8 K but to a higher temperature.} 
Note that the remaining peaks fit well to the experiment already at 8 K and change only marginally increasing the temperature to 30 K.

% B, C, D Peaks
By analyzing the vibrational eigenvectors we can see that the modes B and C correspond to a radial atomic motion, where the atoms are out of phase, while peak D corresponds to the coherent acoustic mode (or breathing mode) where the atoms are in phase.
The fact that we obtain a good agreement between the energetic position of peak D, despite the differences in radii (experimental QD with 18~\AA~ radius and our QD with 11.6~\AA~ radius) suggests a weak confinement effect on the vibrational frequency at these radii.
%
% Peak E
Our calculations show that peak E corresponds to acoustic-like vibrations with radial character, much like the peaks B and C.
%
 % Peak F
In contrast, peak F originates from the LO phonon mode of the CdSe bulk counterpart. We obtain a slightly blue shifted position of this peak, compared to experiment that can be explained by the bond length reduction we obtain at the surface of our CdSe NCs, which does not represent the experimental core/shell situation.
%
% Peak I H G
The peaks identified as I,H,G in the experiment are related to vibrations of the CdS and ZnS shell. These cannot be present in our calculations since we do not use any shell material. The peaks we obtain at around 40~meV are the result of blue shifted CdSe vibrations induced by the light masses of our surface passivants.

\section{High class transition in diamondoids}\label{sec:HighClass}

Until now, we have discussed NCs that exhibit a moderate electron-vibron coupling ($S < 1$) and hence the PLS was strongly dominated by class 1 transitions with a low number of vibrons (mainly $n_g^1 = 1$). In this section, we focus on diamondoids, which are hydrogen-terminated diamond nanoparticles composed of an $sp^3$ hybridized carbon framework\cite{dahl03,schwertfeger08,dahl10} and have strong electron-vibron coupling (large HR factors).  In these structures higher class transitions are expected to become relevant as well as transitions involving $n_g^1 > 1$. The size and symmetry of diamondoids are determined by the number and the geometry of the $sp^3$ hybridized diamond cages. Due to their unique physical properties with simple and well-defined atomic structures, a lot of attentions has been paid to these structures from both experimental and theoretical sides\cite{dahl03,schwertfeger08,dahl10,voros12,drummond05,patrick13,han16}.
Recently, laser-excited fluorescence spectra of free diamondoid molecules were reported by Richter \emph{et~al.} with \textit{ab initio} DFT and time-dependent density functional theory (TDDFT) calculations\cite{richter15}. In their work, high class vibronic transitions, which correspond to the excitation of vibrons of different modes simultaneously, were considered to explain the measured emission spectra.

\begin{center}
\begin{figure*}[t]
\includegraphics[width=0.95\textwidth]{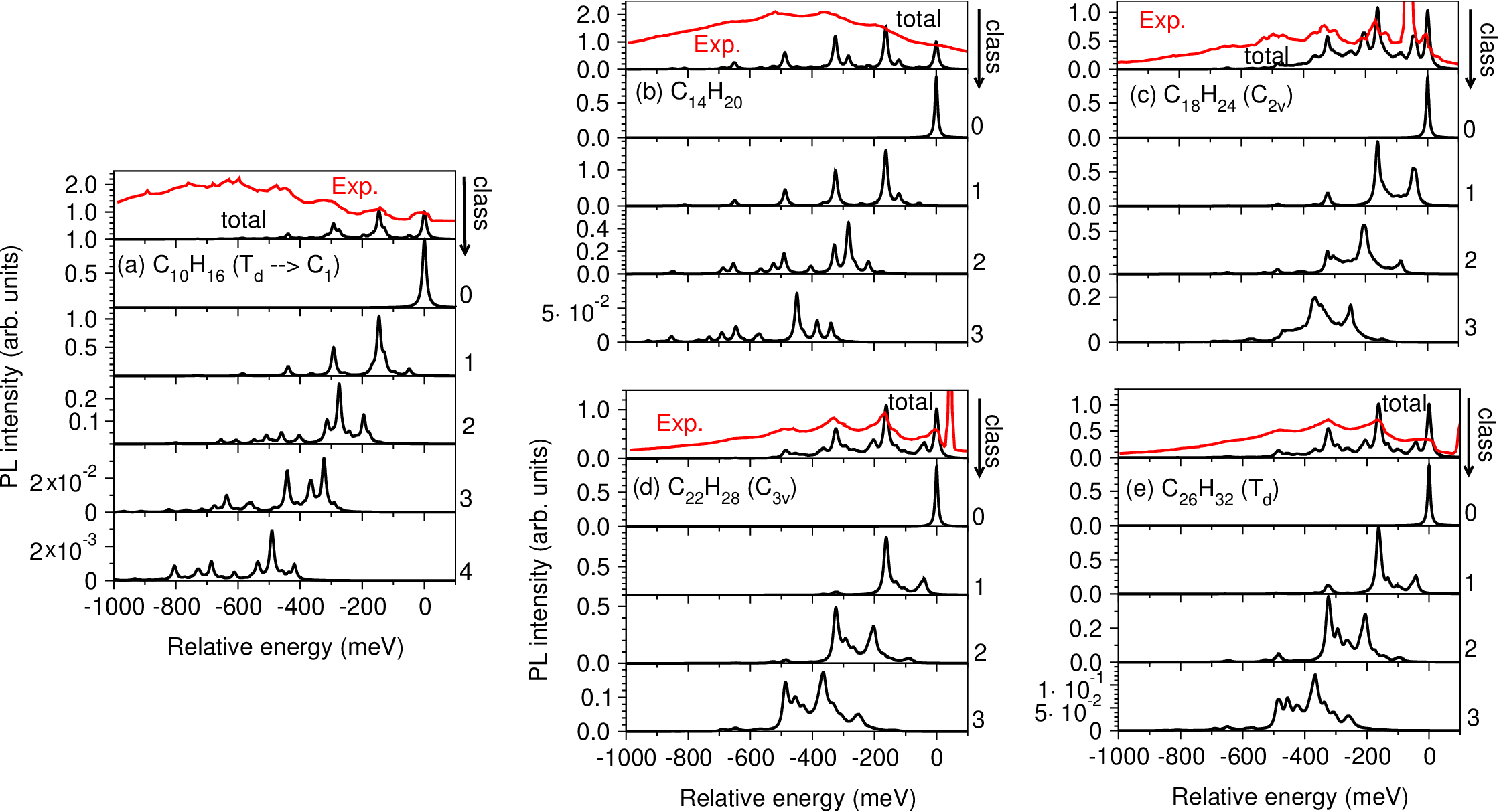}
\caption{\label{fig:highclass_diamondoid}
	Calculated PLS of (a) C$_{10}$H$_{16}$ ($T_d$), (b) C$_{14}$H$_{20}$ ($D_{3d}$), (c) C$_{18}$H$_{24}$ ($C_{2v}$), (d) C$_{22}$H$_{28}$ ($C_{3v}$) and (e) C$_{26}$H$_{32}$ ($T_d$) diamonoids decomposed into contributions from different classes of transitions (see Fig.~\ref{fig:PL_classes} for an explanation of the classes). Class 0 represents the ZPL with the transition from $n_e=0$ to $n_g=0$. The PLS of C$_{10}$H$_{16}$ is calculated until class 4, while (b)-(e) are calculated until class 3. Experimental data taken from Ref.~[\onlinecite{richter15}] are given as red lines in the top panels of (a) and (b). }
\end{figure*}
\end{center}

To calculate the overlap of nuclear wave functions $\langle \phi_m|\phi_n \rangle$ for high class transitions, Richter \emph{et.~al.} \cite{richter15} performed a Duschinsky transformation between the vibrational modes of the ground and excited electronic states and used \textit{ab initio} DFT and TDDFT \cite{richter15,santoro07,barone09}. To avoid the large required computational time and memory usage associated with the Duschinsky transformation approach, we have used a simplified approach introduced in the methods section. 

For the emission at zero temperature ($\ket{\phi_0}$, i.e. all $n_e^i = 0$) we write generally:
\begin{equation}
\label{eq:Iem3_gen}
A_{em}(E)=\sum_{n_g^i}
\left(
\prod_{i=1}^{3N-6}\text{e}^{-S_i}\frac{S_i^{n_g^i}}{n_g^i!} 
\right)
\delta[E_0-E-\sum_i n_g^i \hbar\omega_i] .
\end{equation}
Separating the many terms into classes allows a formulation without the compact index $n_g^i$ used under the sum, so that we can rewrite Eq.~(\ref{eq:Iem3_gen}) as:
%\begin{widetext}
%\begin{eqnarray}
%\label{eq:ne0_2}
%	& A_{em}(E) &= \text{e}^{-S}[ \underbrace{\delta(E_0-E)}_{\text{ZPL}} +
%	\underbrace{\sum_{\alpha}^{n_{\text{max}}}\sum_{i=1}^{3N-6}\frac{S_{i}^{\alpha}}{\alpha!}\delta(E_0-E-\alpha\hbar\omega_i)}_{\text{class 1}}+
%         \underbrace{\sum_{\alpha,\beta}^{n_{\text{max}}}\sum_{\substack{i \;j \\ i\neq j}}^{3N-6}\frac{S_{i}^{\alpha}S_{j}^{\beta}}{\alpha!\beta!}\delta(E_0-E-\alpha\hbar\omega_i-\beta\hbar\omega_j)}_{\text{class 2}}\nonumber \\
%         &&+ 
%         \underbrace{\sum_{\alpha,\beta,\gamma}^{n_{\text{max}}}\sum_{\substack{i \;j \; k\\ i\neq j \neq k}}^{3N-6}\frac{S_{i}^{\alpha}S_{j}^{\beta}S_{k}^{\gamma}}{\alpha!\beta!\gamma!}\nonumber \delta(E_0-E-\alpha\hbar\omega_i-\beta\hbar\omega_j-\gamma\hbar\omega_k)}_{\text{class 3}}+...] 
%\end{eqnarray}
%\end{widetext}

\begin{eqnarray}
\label{eq:ne0_2}
 && A_{em}(E) = \text{e}^{-S}[ \underbrace{\delta(E_0-E)}_{\text{ZPL}} + \nonumber \\
&&\underbrace{\sum_{\alpha}^{n_{\text{max}}}\sum_{i=1}^{3N-6}\frac{S_{i}^{\alpha}}{\alpha!}\delta(E_0-E-\alpha\hbar\omega_i)}_{\text{class 1}} + \nonumber \\
&&
\underbrace{\sum_{\alpha,\beta}^{n_{\text{max}}}\sum_{\substack{i \;j \\ i\neq j}}^{3N-6}\frac{S_{i}^{\alpha}S_{j}^{\beta}}{\alpha!\beta!}\delta(E_0-E-\alpha\hbar\omega_i-\beta\hbar\omega_j)}_{\text{class 2}} + \nonumber \\ 
&& 
 \underbrace{\sum_{\alpha,\beta,\gamma}^{n_{\text{max}}}\sum_{\substack{i \;j \; k\\ i\neq j \neq k}}^{3N-6}\frac{S_{i}^{\alpha}S_{j}^{\beta}S_{k}^{\gamma}}{\alpha!\beta!\gamma!}\nonumber \delta(E_0-E-\alpha\hbar\omega_i-\beta\hbar\omega_j-\gamma\hbar\omega_k)}_{\text{class 3}} \nonumber \\
 &+& ...] 
\end{eqnarray}

where,  $S=\sum_{i=1}^{3N-6}S_i$ is the total HR factor, $i$, $j$, and $k$ denote different vibrational modes and $\alpha$, $\beta$, and $\gamma$ denote the phonon (vibron) quantum numbers. For practical purposes the sums will be truncated at the certain maximum number of vibrons $n_{\text{max}}$.

In Fig.~\ref{fig:highclass_diamondoid} (a), we present the PLS of C$_{10}$H$_{16}$ with $T_{d}$ symmetry (adamantane) resolved by classes, and compare the results to experiment \cite{richter15} (red line).
The vibrational modes with dominant PL intensity for class 1 transitions are breathing type modes in agreement with our earlier work \cite{han19} where we showed that modes with $\Gamma_1$ point group symmetry have nonvanishing HR factors.
Here we want to highlight the importance of the symmetry of the excited states. Indeed, while the  point group symmetry of the ground state is $T_d$, the electronically excited state has a reduced symmetry, as described earlier \cite{richter14,banerjee14}, and we perform our CDFT calculation without imposing any symmetry constraints. 
In Fig.~\ref{fig:highclass_diamondoid} (b)-\ref{fig:highclass_diamondoid}(e), we show the experimental (from Ref.~[\onlinecite{richter15}]) and theoretical results for (a) C$_{14}$H$_{20}$ with $D_{3d}$ symmetry (diamantane), (b) C$_{18}$H$_{24}$ with $C_{2v}$ symmetry (triamantane), (c) C$_{22}$H$_{28}$ with $C_{3v}$ symmetry ([1(2)3]tetramantane) and (d) C$_{26}$H$_{32}$ with $T_d$ symmetry ([1(2,3)4]pentamantane) diamondoids.  
{The comparison with experiment shows a very good agreement for the diamondoids we studied except for C$_{14}$H$_{20}$.  Note that the band gaps of C$_{10}$H$_{16}$ and C$_{14}$H$_{20}$ are beyond the photon energy range of the optical parametric oscillator laser used in the experiment. Instead, synchrotron radiation was used as the pump source. In contrast to C$_{10}$H$_{16}$, the excitation energy of the synchrotron radiation significantly exceeds the band gap of C$_{14}$H$_{20}$. This results in additional intramolecular vibrational redistribution and causes the experimental spectrum to be considerably congested. In addition, a lower signal-to-noise ratio in synchrotron measurement leads to the relatively strong background in the measured PLS of the C$_{14}$H$_{20}$ cluster. See Ref.~\onlinecite{Richter_thesis} for a detailed discussion.}
In Fig.~\ref{fig:highclass_diamondoid} we also generally notice that the spectrum intensity has relatively strong contributions from class 2 and even class 3 transitions [class 4 transitions are only relevant in C$_{10}$H$_{16}$; see Fig.~\ref{fig:highclass_diamondoid} (a)].
{A further benchmark of the quality of our high class model is given by the direct comparison to the more expensive Duschinsky rotation approach~\cite{richter15,Richter_thesis}. We obtain a very good agreement between the line shapes and peak positions for all structures, including the C$_{14}$H$_{20}$.}

\begin{center}
\begin{figure}%[h]
\includegraphics[width=\columnwidth]{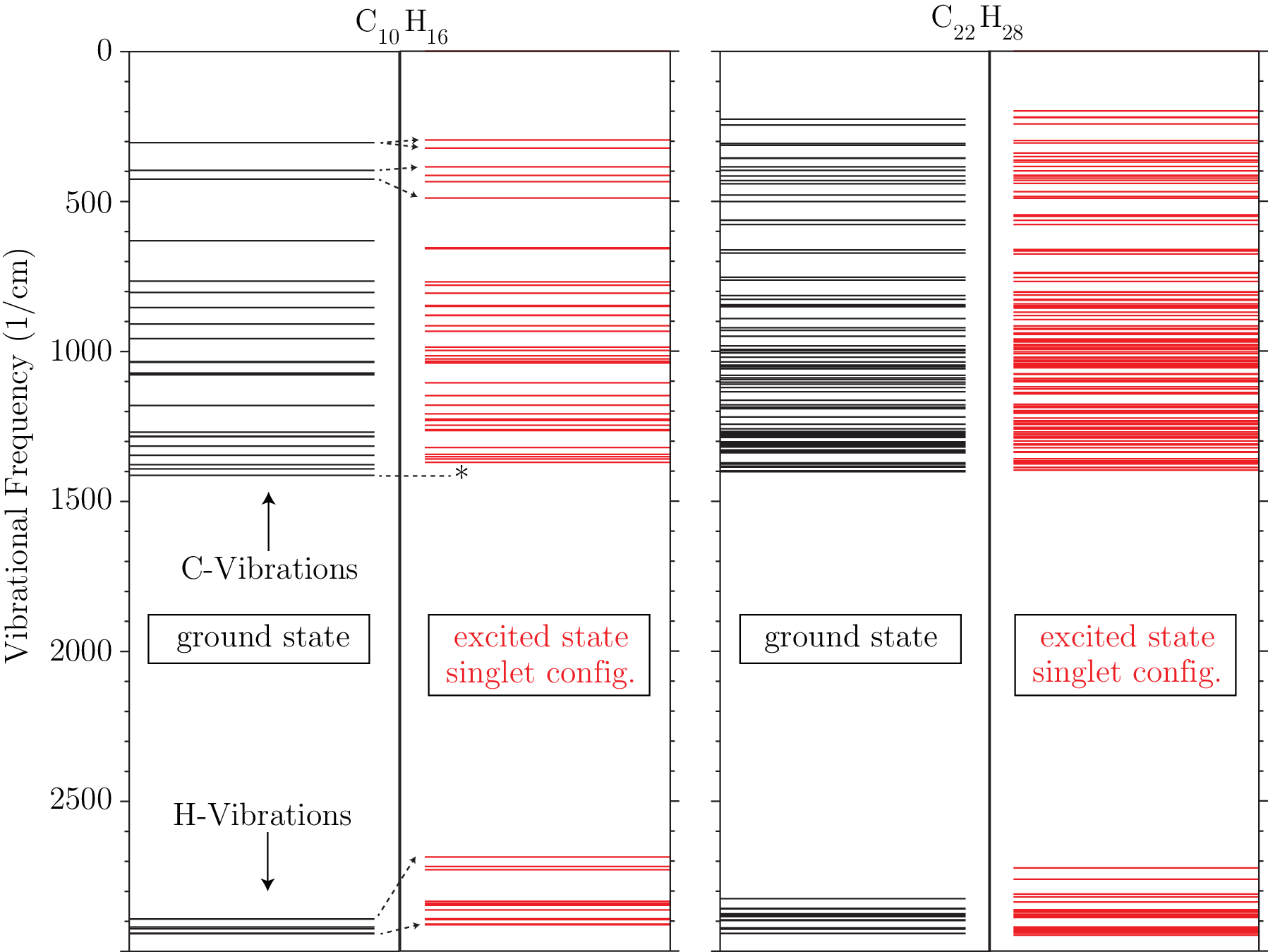}
\caption{\label{fig:vib_freq_compare_C10_C22}
Calculated vibrational frequencies in the electronic ground and excited states for  C$_{10}$H$_{16}$ and C$_{22}$H$_{28}$.}
\end{figure}
\end{center}

As mentioned in Sec.~\ref{sec:theory} we do not take into account the change in frequency going from the ground to the excited state. While this approach is well justified for larger structures it represents an approximation for smaller structures such as small diamondoids. In Fig.~\ref{fig:vib_freq_compare_C10_C22} we show the vibrational frequencies obtained for the ground electronic state and for the singlet excited electronic state for C$_{10}$H$_{16}$ and C$_{22}$H$_{28}$. We can see for C$_{10}$H$_{16}$ a small red shift of the C-vibrational modes (indicated by an asterisk *) and a large red shift for the H vibrations, which can be explained by the fact that upon excitation an electron is promoted from a bonding to an antibonding orbital causing a weakening of the bonds. We can also observe at low frequencies and for the H vibrations that degenerate vibrational modes split (indicated by small arrows), due to the reduced symmetry of the excited state. Already for the slightly larger C$_{22}$H$_{28}$ structure, the vibrational frequency change is less severe.

\section{Summary}
In summary, we present a theoretical approach which allows us to include vibronic coupling effects in the treatment of the optical properties of NCs. While the method relies on a few approximations and is computationally very advantageous, it is demonstrated to agree well with experiment and  with high-level \textit{ab initio} approaches (for very small structure, where these approaches are feasible).  

With this approach we study vibronic coupling effects and the emergence of phonon satellites in Si and CdSe colloidal NCs with radii ranging from 7.8 to 14.9~\AA~ at both zero and finite temperatures. We further study multiphonon processes  (high class vibronic transitions) using five different small carbon structures called diamondoids. 
We find the following. (i) The phonon satellites in the PLS of colloidal CdSe and Si semiconductor NCs are dominated by  vibrations with breathing character: the coherent acoustic mode (all atoms in phase) and breathing modes where core and shell atoms have opposite phases. Contributions from optical-like phonon modes are strong in CdSe, especially in larger structures, while it is much weaker, but still observable, in Si NCs.  
(ii) Phonon satellites at zero temperature are mainly induced by vibronic transitions between the excited electronic state with phonon number $n_e=0$ to the ground electronic state with phonon number $n_g=1$,  as illustrated in Fig.~\ref{fig:FrankCondon}. This observation is in line with the small HR factors of our NCs. Higher order transition between $n_e$ = 0 and $n_g$ = 2 or 3 can only be seen in NCs  with stronger exciton-phonon coupling. (iii) For PL at finite temperatures, the excited electronic state shows as thermal vibron occupation ($n_e>0$ occupied) and the transition to the vibronic ground state (ground electronic state with $n_g=0$) leads to significant anti-Stokes PL peaks starting at around 140K. These thermally excited vibronic states along with the reduction of phonon lifetime (not calculated in this work) deeply alters the line shape of PLS at high temperatures.
(iv) We obtain significant contributions from high class transitions for our small diamondoid structures. In these structures, the effect is very strong, up to class 4 for our smallest adamantane structure, and neglecting them would lead to significant deviations from experiment. For the larger Si and CdSe NC these effects are small. Since high class transitions are stronger in structures with large HR factors, we expect them to become relevant for some defects.

Our general good agreement with experiment allows us to interpret experimental data in a peak-by-peak fashion. We expect that this ability to accurately simulate the PLS, including vibronic coupling, will be beneficial to the interpretation of low temperature PL measurements in the future.

\begin{acknowledgments}
We thank Dr. R. Richter for helpful discussions and providing us with the raw experimental data. P.H. acknowledges financial support by the National Natural Science Foundation of China under Grants No. 12174270 and No. 11774243. P.H. and G.B. acknowledge financial support by the Chinesisch-Deutsches Mobilit\"atsprogramm von Chinesisch-Deutsches Zentrum f\"ur Wissenschaftsf\"orderung under Grant No. M-0225. This work is supported by the DFG project GZ: BE 4292/4-1 AOBJ: 651735 ``Resonant Raman spectroscopy as tool to investigate colloidal semiconductor nanocrystals''. Most of the simulations were performed on the Cray XC40 Hazel Hen Supercomputer Cluster at the High Performance Computing Center Stuttgart (HLRS). 
\end{acknowledgments}

\end{document}